\documentclass[aps, prappl,superscriptaddress,twocolumn,showpacssuperscriptaddress%
phys,
jmp,
amsmath,amssymb,
reprint,
author-numerical,%
sort,
]{revtex4-1}

\usepackage{graphicx}
\usepackage{dcolumn}
\usepackage{bm}
\usepackage{multirow}
\usepackage{url}
\usepackage{epsfig,rotating,graphicx,color}
\usepackage{natbib}

\begin{document}


\title{Simulation and measurement of stray fields for the manipulation of spin-qubits in one- and two-dimensional arrays}

\affiliation{%
 IBM Research-Zurich, Säumerstrasse 4, 8803 Rüschlikon, Switzerland}%
\affiliation{%
 Department of Physics, University of Basel, 4056 Basel, Switzerland}%
\author{Michele Aldeghi}
\author{Rolf Allenspach}
\affiliation{%
 IBM Research-Zurich, Säumerstrasse 4, 8803 Rüschlikon, Switzerland}%
\author{Andriani Vervelaki}%
\author{Daniel Jetter}%
\author{Kousik Bagani}%
\author{Floris Braakman}%
\affiliation{%
 Department of Physics, University of Basel, 4056 Basel, Switzerland}%
\author{Martino Poggio}%
\affiliation{%
 Department of Physics, University of Basel, 4056 Basel, Switzerland}%
\affiliation{%
 Swiss Nanoscience Institute, University of
Basel, 4056 Basel, Switzerland}%
\author{Gian Salis}%
\email{gsa@zurich.ibm.com}
\affiliation{%
 IBM Research-Zurich, Säumerstrasse 4, 8803 Rüschlikon, Switzerland}%

\date{\today}

\begin{abstract}
The inhomogeneous magnetic stray field of micromagnets has been extensively used to manipulate electron spin qubits. By means of micromagnetic simulations and scanning superconducting quantum interference device microscopy, we show that the polycrystallinity of the magnet and non-uniform magnetization significantly impact the stray field and corresponding qubit properties. We find that the random orientation of the crystal axis in polycrystalline Co magnets alters the qubit frequencies by up to 0.5 GHz, compromising single qubit addressability and single gate fidelities. We map the stray field of Fe micromagnets with an applied magnetic field of up to 500 mT (mimicking conditions when operating qubits), finding field gradients above 1 mT/nm. The measured gradients and the lower magnetocrystalline anisotropy of Fe demonstrate the advantage of using Fe instead of Co for magnets in spin qubit devices. These properties of Fe also enabled us to design a 2D arrangement of nanomagnets for driving spin qubits distributed on a 2D lattice.
\end{abstract}

\maketitle

Engineering the stray field profile of magnets at the nanoscale finds widespread applications, ranging from technological applications for magnetic storage~\cite{tape} and sensing~\cite{tuma2014high} to studying complex phenomena like phase transitions~\cite{skjaervo2020advances} and nanoscale imaging~\cite{marchiori2022nanoscale}. One application that requires exquisite control of the stray field of nanomagnets is quantum computation with electron spins. There, spins of electrons electrostatically confined in quantum dots~\cite{Lossdivincenzo} define qubits that can be manipulated via electron dipole spin resonance (EDSR)~\cite{allenspach2008method, tokura2006coherent, Pioro-Ladriere2008}, in which the electron's wavefunction is resonantly displaced  within the inhomogeneous stray field of a micromagnet. Single qubit gate fidelities of 99.9\% \cite{yoneda2018quantum} and control of up to six qubits \cite{philips2022universal} were achieved, with two-qubit gates fidelity reaching 99.5\% \cite{noiri2022fast}.

Qubit gate fidelity and addressability depend on the interplay between the electron confinement (controlled by electric fields) and the local magnetic field profile given by the magnet. The shape of the magnet needs to be adapted to the device-specific qubit arrangement and number~\cite{Yoneda_robust}, with the goal to maximize qubit gate fidelity while providing addressability of individual qubits. Gradients of field components transversal to the main field direction are used for Rabi driving of the qubits by an AC electric field, while longitudinal gradients spectrally separate neighboring qubits to enable individual addressability. However, in combination with charge noise, the longitudinal field gradient will induce qubit dephasing~\cite{struck2020low}. Making informed decisions about the design of the magnet therefore requires an exact knowledge of the stray field profile~\cite{neumann2015simulation}.

This letter shows that non-uniform magnetization, polycrystallinity and fabrication imperfections of the lithographically defined magnets affect qubit properties to a much larger extent than previously assumed. We measure the stray field of Fe magnets by scanning superconducting quantum interference device (SQUID) microscopy (SSM)~\cite{wyss2022magnetic} and compare it with micromagnetic simulations, disentangling the effects of polycrystallinity, surface roughness and reduced saturation magnetization. We then apply our model to a spin-qubit experiment with the same magnet design~\cite{philips2022universal}, showing that spin qubit properties can be accurately modeled. Finally, we propose a magnet geometry resilient to polycristallinity-induced stray-field variations that enables manipulation of spin qubits arranged on a two-dimensional lattice.



\begin{figure}[ht]
	\centering
	\includegraphics[width=0.48\textwidth]{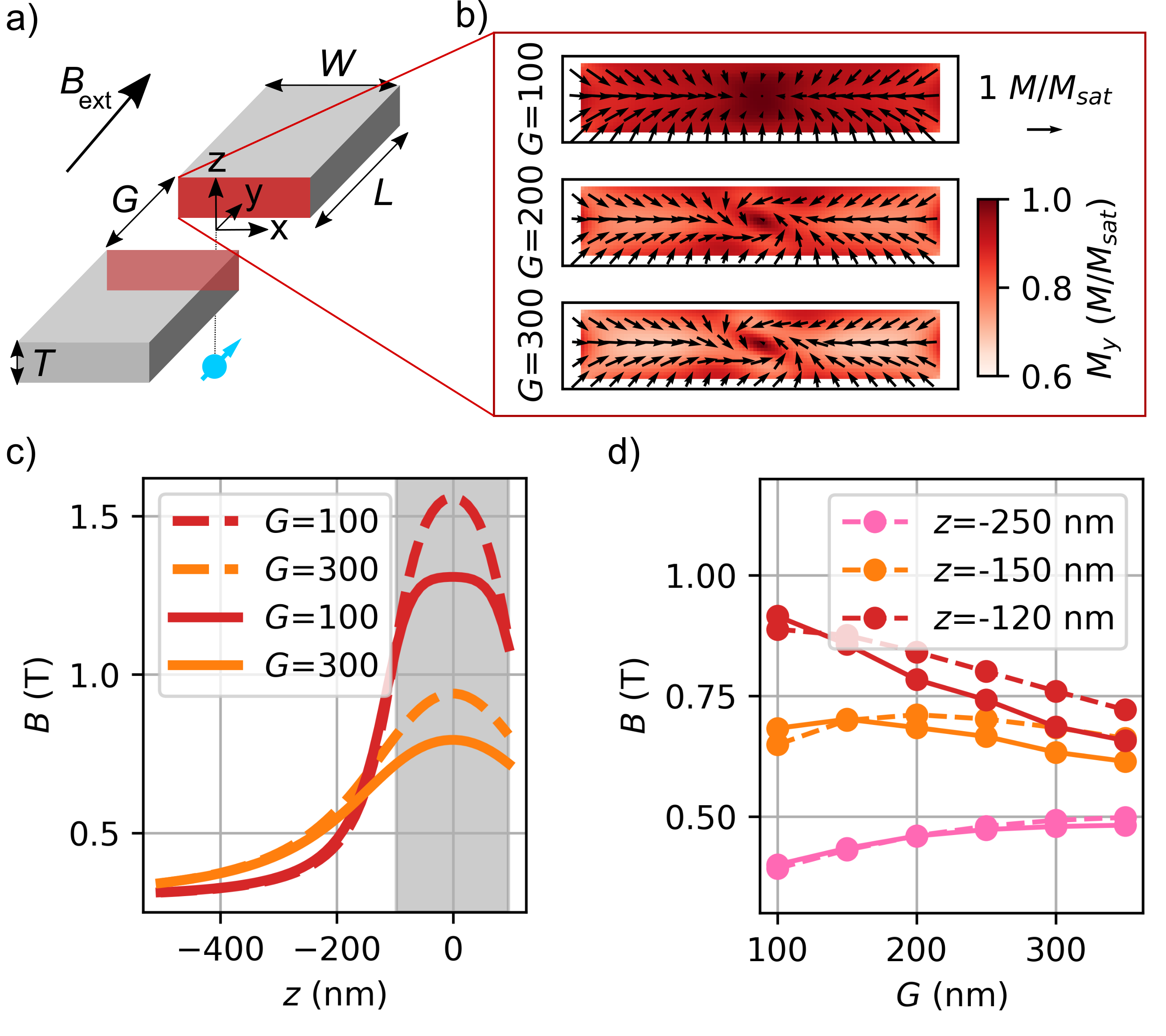}
	\caption{\label{fig:fig_1} Comparison of magnetization pattern and magnetic field ($B$) between the MS approximation and EM micromagnetic simulations. (a) Device geometry, showing magnets (grey), qubit (blue spins) and external field arrangement. The origin of the coordinate system is centered in the middle of the gap. (b) The magnetization pattern (EM) at the surface of the magnet depends on the gap distance $G$ (given in nm) with ($B_{\mathrm{ext}}=0.3$\,T). In-plane components are shown as arrows, the out-of-plane component as color. The length of the arrow above the colorscale corresponds to saturation magnetization. (c, d) Stray field magnitude for different height and gap distances (EM continuous lines, MS dashed lines). The grey shaded area in (c) shows the $z$ position of the magnets.}
\end{figure}

Two methods are commonly used to compute the stray field from micromagnets: one based on the macrospin approximation (abbreviated with "MS") and one where the magnetization pattern is computed by micromagnetic simulations to reach an energy minimum (abbreviated with "EM"). In the MS approach, the magnetization within the entire magnet is assumed to be uniform. In the EM approach, a magnetization pattern is calculated by minimizing the overall energy of the system~\cite{finitediff}. The magnetic field ($B$) calculated from the magnetization pattern differs for the two methods, and therefore different qubit properties are predicted. We note that the computationally lighter MS approximation is commonly used for the design of micromagnets~\cite{Brunner11, Yoneda_robust, philips2022universal, Pioro-Ladriere2008}, even though complete magnetic saturation is not achieved at external magnetic fields usually applied in spin qubit experiments (ranging from 0.2 to 1\,T). 

We first show that the MS approximation cannot accurately model the stray field for the manipulation of spin qubit. For this, we illustrate the differences between the MS and the EM results on the simplest magnet geometry useful for EDSR driving, where the opposite poles of two magnets are brought close together [Fig.~\ref{fig:fig_1}(a)], generating a large stray field that can be tuned by changing the gap distance $G$ between the two magnets. In our case, the magnets are cuboids (width $W= 1000$\,nm, length $L= 3000$\,nm, thickness $T= 200$\,nm). We set the origin of the coordinate system in the center of the gap, with an external magnetic field $B_{\mathrm{ext}}=0.3$\,T applied along the $y$ direction and the magnets in the $x-y$ plane. The position of the qubits is assumed to be centered in the gap with an offset in the $z$ direction. In this region, the stray field is mostly set by the magnetization patterns on the two opposing surfaces facing the gap [highlighted in red in Fig.~\ref{fig:fig_1}(a)]. 

In the MS approximation, the magnetization points perpendicularly to these surfaces, creating a large stray field within the volume $V_{\textrm{gap}}$ spanned by the two surfaces. At small fields, such a magnetization pattern costs more magnetostatic energy than what is saved in Zeeman and exchange energy. Indeed, the EM simulation predicts a rotation of the magnetization from the out-of-plane towards the in-plane direction at these two surfaces, leading to a stray field reduction. Increasing $G$ leads to more pronounced rotations, due to the reduced magnetostatic interaction between the two magnets [Fig.~\ref{fig:fig_1}(b)].

We compare $B$ obtained from the MS and EM approaches in Fig.~\ref{fig:fig_1}(c), where we focus on profiles along $z$ taken at $x,y=0$. For both EM and MS simulations, we observe that a larger gap \textit{increases} $B$, as shown in Fig.~\ref{fig:fig_1}(d) for $z=-250$\,nm, which is a consequence of the dipolar nature of the field distribution close to the gap. The optimal gap size to maximize the stray field therefore depends also on $z$. The magnetization rotation accounted for in the EM simulation leads to a change of the Larmor frequency $f_{\mathrm{L}}= \frac{g\mu_B}{h}B$ on the order of \textit{several} GHz if compared to the MS approximation (e.g. for $z=-200$\,nm and $G=300$\,nm, the difference is 65 mT which corresponds to a qubit frequency difference of $\Delta f_{\mathrm{L}}=1.82$\,GHz, with $\mu_{\mathrm{B}}$ the Bohr's magneton, $h$ Planck's constant and a g-factor $g=2$). This is much larger than the typically targeted $\Delta f_{\mathrm{L}}$ between neighboring qubit frequencies. We note that these differences need to be carefully engineered in a larger array of spin qubits to avoid crosstalk of the EDSR drive with close-by qubits~\cite{Yoneda_robust, Brunner11} (see also SI), emphasizing the importance of a precise knowledge of the expected stray field.

\begin{figure*}[ht]
	\centering
	\includegraphics[width=1\textwidth]{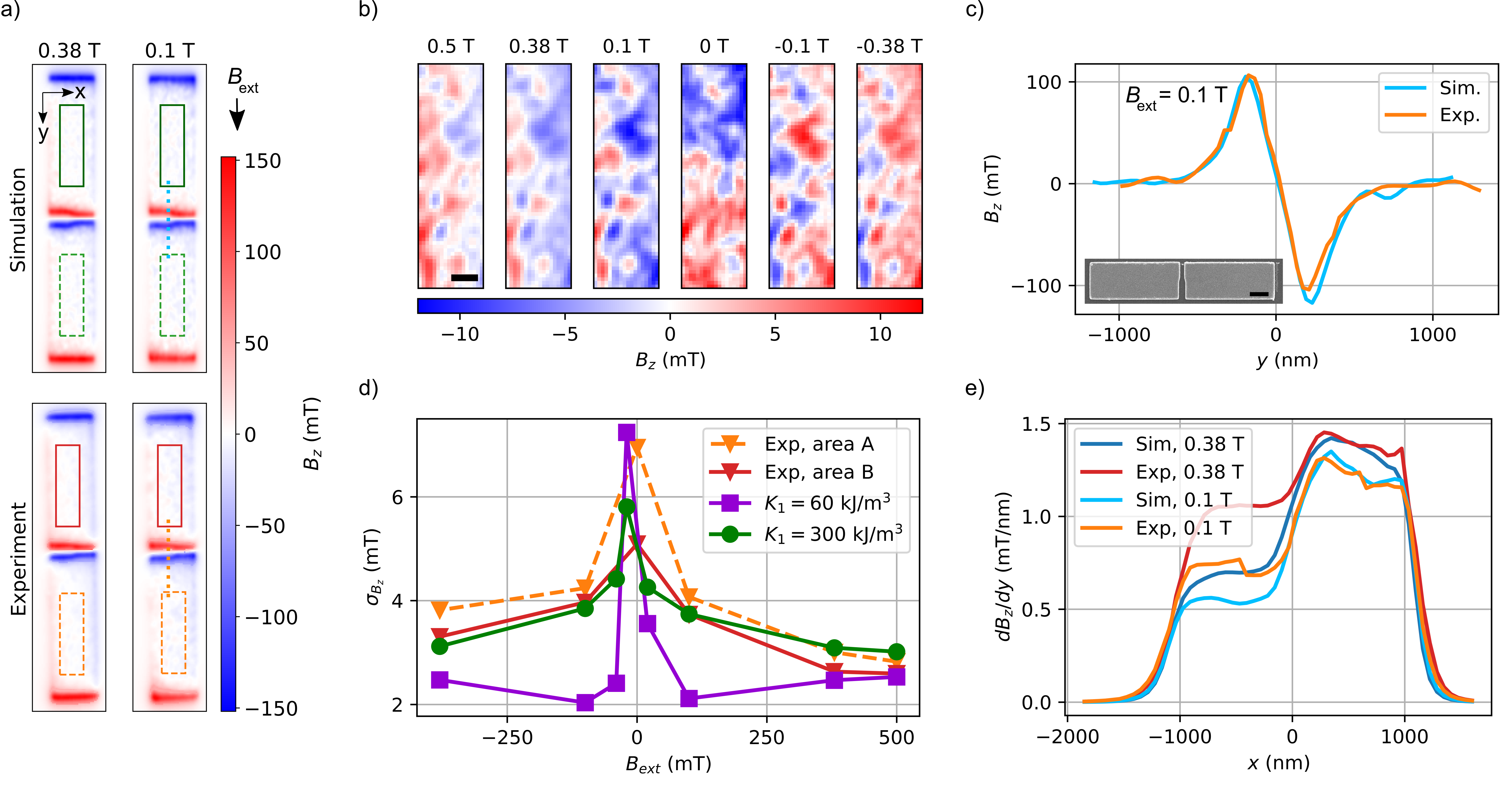}
	\caption{\label{fig:fig_2} SSM measurements and simulations of the stray field at $z=70$\,nm (where $z=0$ is at the upper surface of the magnets). (a) $B_{\mathrm{z}}$ stray field comparison between micromagnetic simulations and SSM measurements for different external fields. (b) Magnification of the area framed by the orange dashed line in (a) for different $B_{\textrm{ext}}$. The scale bar is 500 nm. (c) Simulated and measured $B_{\mathrm{z}}$ along the dotted line in (a). The inset shows an SEM image of the measured magnets. The scale bar is 1000 nm. (d) Standard deviation of $B_{\mathrm{z}}$ versus $B_{\textrm{ext}}$ for the regions enclosed by the rectangles in (a). The lines are guides to the eye. (e) Fitted driving gradient $dB_z/dy$ in the middle of the gap ($y=0$) between the magnets at two external fields.}
\end{figure*}

Up to here, we have considered ideal magnet geometries with uniform magnetic properties throughout the magnet volume. Now, we take into account lithographically induced modifications of the shape and the polycrystalline structure of nanofabricated EDSR magnets. Rounding effects at the edges of the magnet modify the magnetization pattern, with variations of $B$ of up to several GHz (e.g. the $\Delta f_{\mathrm{L}}$ between a straight and a semicircle profile at $z=-120$\,nm is 4.08\,GHz, see SI). Also surface roughness and sample polycrystallinity strongly affect the stray field~\cite{meyer2003scanning}, both of which are typically present in nanofabricated EDSR magnets. Surface roughness varies the stray field through magnetization rearrangement at the corrugated surface of the magnet and through modulation of the effective distance between the magnet edge and the qubit array. Polycrystallinity further modulates the stray field if the material exhibits a magnetocrystalline anisotropy (MCA), since the magnetization direction will then be different for each crystallite. 

\begin{figure*}[ht]
	\centering
	\includegraphics[width=1\textwidth]{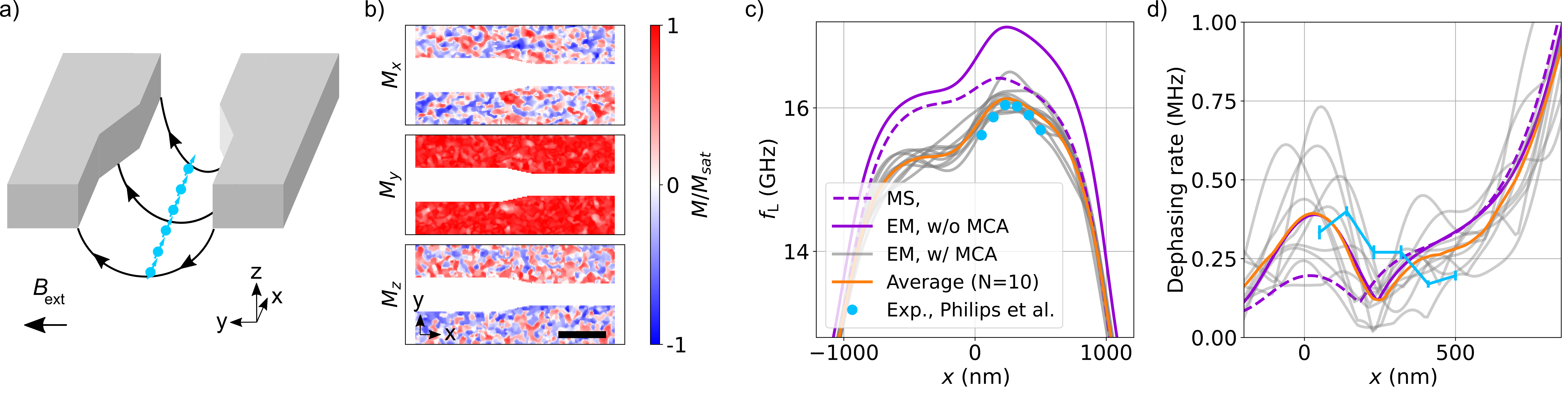}
	\caption{\label{fig:fig_3} (a) Device geometry, adapted from~\cite{philips2022universal}. (b) Magnetization pattern of one out of the 10 different crystallite patterns used in the EM simulations at $B_{\mathrm{ext}}=380$\,mT. The scale bar is 500 nm. (c) $f_{\mathrm{L}}$ and (d) dephasing rate profile along the line of qubit for the MS and EM with and without MCA simulations. The 10 different gray lines correspond to 10 different crystallite patterns, and their average is shown in orange. The light blue points in (c) and (d) are the respective values obtained from qubit measurements by Philips et al.~\cite{philips2022universal}.}
\end{figure*}

We investigate the effect of these non-idealities on $B$ following the design used by Philips et al. \cite{philips2022universal} [sketched in Fig.~\ref{fig:fig_3}(a)]. E-beam evaporation and lift-off have been used to pattern $100$\,nm thick Fe magnets capped by $5$\,nm of Pt on a Si wafer with native oxide [inset Fig.~\ref{fig:fig_2}(c)]. We use SSM to image $B_z(x,y)$ at 70 nm above the magnet surface at 4.2\,K, with $B_{\mathrm{ext}}$ applied along $y$. We note that the stray field significantly exceeds the SQUID monotonic response range of about 100\,mT. Thus, we employ a method to extract field values from a modulating SQUID signal as explained in the SI. Fig.~\ref{fig:fig_2}(a) shows the large $B_z$ at the magnet ends, as is expected from the $B_{\mathrm{ext}}$ direction. We compare these measurements with EM simulations where polycrystalline magnets with a mean isotropic crystallite size of 40 nm and a corrugated surface with a maximum height variation of 20 nm were assumed, consistent with atomic force microscopy (AFM), x-ray diffraction (XRD) and scanning electron microscopy (SEM) measurements (see SI). We mimic the finite size of the scanning SQUID loop in the simulation by applying a square filter with an edge size of 110 nm.
The simulations and the experiment are in excellent agreement what concerns the large field at the two magnetic pole pairs. The inner two magnetic poles provide the stray field suitable for EDSR drive of spin qubits. We notice some small modulation of $B_z$ over the entire magnet structure. To better understand the origin of this, we focus on the regions between the poles of each magnet [framed in Fig. 2(a)]. There, $B_z$ fluctuates with a peak-to-peak amplitude of 10\,mT [Fig.~\ref{fig:fig_2}(b)], and the overall pattern is inverted if the external field direction is reversed. The amplitude of these modulations decreases with increasing fields, as seen quantitatively from a plot of the standard deviation of the measured field ($\sigma_{B_z}$) versus $B_{\mathrm{ext}}$ in Fig.~\ref{fig:fig_2}(d).

In Fig.~\ref{fig:fig_2}(d) we compare the SSM experiment with EM simulations. The experimental $\sigma_{B_z}$ curves show a peak at $B_{\mathrm{ext}}$ close to the coercive field ($\approx 20$\,mT). If $B_{\mathrm{ext}}$ is increased, $\sigma_{B_z}$ monotonically decreases. We can reproduce the experimental data assuming either a cubic MCA constant of $K_1 = 300$\,kJ/m$^3$ and 40 nm crystallite mean size ($\lambda$) or $K_1 = 60$\,kJ/m$^3$ and $\lambda=120$\,nm (data not shown). For $K_1 = 60$\,kJ/m$^3$ and $\lambda=40$\,nm, $\sigma_{B_z}$ initially decreases to a lower level but then monotonically increases from $B_{\mathrm{ext}}=100$\,mT, reaching the same value as in the experiment only at $B_{\mathrm{ext}}=500$\,mT.

We can understand the different trends by inspecting the simulated magnetization direction at different $B_{\mathrm{ext}}$, from which we can draw conclusions about the relative size of the Zeeman, MCA and magnetostatic energies. At $\lvert B_{\mathrm{ext}} \rvert \approx 20$\, mT the magnetization state is dominated by local defects and MCA, leading to a peak in $\sigma_{B_z}$. At $B_{\mathrm{ext}}= 500 $\,mT the Zeeman energy dominates and the magnetization is mostly aligned to the direction of the magnetic field, such that surface roughness determines the value of $\sigma_{B_z}$ (due to the formation of magnetic charges on the corrugated surface). At intermediate fields ($100 < \lvert B_{\mathrm{ext}} \rvert <380$\,mT), the ratio between the MCA and the magnetostatic energy determines the trend of $\sigma_{B_z}$ with decreasing $B_{\mathrm{ext}}$. For small MCA, the local magnetization progressively orients more parallel to the surface to minimize the amount of magnetic charges along the corrugations, resulting in a decrease of $\sigma_{B_z}$. For large MCA, however, the magnetization in each crystallite orients towards the respective easy direction, whereby $\sigma_{B_z}$ increases.

We now discuss why we need to adapt either $\lambda$ to 120 nm (three times larger than the measured crystallite size) or $K_1$ to 300\,kJ/m$^3$ (five times larger than the literature value) to match the experimental data. On the one hand, the simulation where $K_1$ is increased could account for the elongated shape of the crystallites (XRD and AFM measurements indicate 15 nm along the smallest and 40 nm along the largest axis, see SI). The larger $K_1$ could thus emulate a mixed anisotropy (composed of shape anisotropy and MCA~\cite{cullity2011introduction, nesbitt1954factors}). On the other hand, increasing $\lambda$ could suggest the formation of so-called "interaction-domains"~\cite{Craik1960160, hadjipanayis1999nanophase}. These domains are formed by nanocrystallites that combine together into effective domains with combined magnetization direction. Indeed, exchange-coupled crystallites with size comparable to the exchange length ($l_{\mathrm{ex}}=\sqrt{2 A/(\mu_0 M_{\mathrm{sat}}^2)}\approx5$\,nm, where $A$ is the exchange stiffness constant) cannot vary their magnetization direction abruptly from one crystallite to another~\cite{Goll, schafer1991domain}. The two possibilities exhibit different correlation lengths of the $\sigma_{B_z}$ variations, which we however cannot distinguish in the experiment because of the finite size of the SQUID loop. We also note that the presence of mixed anisotropy and the formation of interaction domains are not mutually exclusive, such that a combination of both effects may take place.

We now draw our attention to the magnetic field within the gap between the two magnets. Excellent matching between the simulations and the SSM experiment at $B_{\mathrm{ext}}=100$\,mT is found [Fig.~\ref{fig:fig_2}(c)]. The driving gradient $dB_z/dy$ is calculated both for the SSM measurement as well as for the simulations by fitting the data with a polynomial function and taking its analytical derivative (SI). The comparison between simulation and experiment is shown in Fig.~\ref{fig:fig_2}(e) (additional external fields in the SI). We report gradients $dB_z/dy$ larger than $1$\,mT/nm, currently the largest stray field gradient measured by SSM. The impact of MCA on the driving gradient is small: simulations with different crystallite patterns deviate by less than 5\%. We note that the actual gradients experienced by the qubits may be even larger (SI), since the finite size of the SQUID loop (110 nm) is considerably larger than the electron wavefunction~\cite{aldeghi2023modular}. This confirms that Fe micromagnets would provide large and reproducible driving gradients for EDSR drive of electron spin qubits.

Despite its smaller saturation magnetization and larger MCA than for Fe, Co is currently the most commonly used material for the fabrication of EDSR magnets~\cite{Pioro-Ladriere2008, watson, struck2020low, yoneda2023noise}). We now investigate the effect of magnetization pattern relaxation, nanofabrication-related imperfections and polycrystallinity on the qubit properties in a recent 6-qubits experiment~\cite{philips2022universal} where Co was used as magnetic material. The design [Fig.~\ref{fig:fig_2}] was intended to provide a linear decrease of $f_{\mathrm{L}}$ for six qubits along the $x$ direction, according to calculations made with the MS approximation and assuming perfectly symmetric and sharp edges~\cite{philips2022universal}. Within these approximations, the linear change of $f_{\mathrm{L}}$ is a consequence of the linear variation of the gap size along $x$. Nevertheless, the experimentally measured $f_{\mathrm{L}}$ show a parabolic dependence on the qubit position along $x$ [Fig.~\ref{fig:fig_3}(c)], which we explain in the following.

Impurity inclusions and surface oxidation degrade the quality of the magnetic material, causing a reduction of the saturation magnetization (see SI). We expect that for unprotected magnets (like the ones used in~\cite{philips2022universal}) the saturation magnetization is reduced, and set $M_{\mathrm{sat}}=1.11$\,MA/m to fit the experimental values (23\% reduction). Also, nanofabrication patterning may lead to a modification of the intended magnet edge profile, as was the case for this experiment. Thus, we take into account the measured triangular profile in our simulation (see SI). We then account for MCA by setting a random crystallite pattern with an average size of 40\,nm, as extracted from SEM images (SI), and a uniaxial MCA constant of $K_{\mathrm{u}}=600$\,kJ/m$^3$ (the literature value for Co~\cite{cullity2011introduction}). The resulting magnetization at $B_{\mathrm{ext}}=380$\,mT for one specific crystallite pattern is shown in Fig.~\ref{fig:fig_3}(b), highlighting the magnetization direction modulation. 

In Fig.~\ref{fig:fig_3}(c) EM simulations with 10 different crystallite patterns are compared with an EM simulation without MCA, a MS simulation and the Philips et al. experiment~\cite{philips2022universal}. The random orientation of the different crystallites effectively acts like a reduction of the overall magnetization, thereby also reducing the stray field at the qubit position. In addition, different crystallite patterns spatially modulate the $f_{\mathrm{L}}$ with an amplitude reaching $0.5$\,GHz [Fig.~\ref{fig:fig_3}(c)]. The overall trend is well captured by the average of the ten crystal patterns, but the variations suggest that samples grown under identical conditions will provide stray fields with significantly different profiles. The EM simulation without MCA shows larger fields than the MS simulation, which we explain by the magnetization rotation along the $z$ direction caused by the triangular profile of the magnets. 

We now use the simulations to estimate the charge noise displacement $\delta$ acting on the qubits, gaining insight into a parameter that is difficult to measure directly. The qubit dephasing rate $\Gamma$ is linked to $\delta$ by $\Gamma = \pi\sqrt{2}\frac{\gamma_e}{h}(\frac{dB_{y}}{dx}+\frac{dB_{y}}{dy})\delta$ (SI). We plot the dephasing rates ($\Gamma := 1/T_2^*$) experimentally measured~\cite{philips2022universal} in Fig.~\ref{fig:fig_3}(d) and use $\Delta$ as the fitting parameter, finding good agreement between the averaged curve and the measured values for a displacement amplitude of $\delta= 109$\,pm. This value is comparable with what has been estimated in other SiGe platforms~\cite{Kawakami2014} but here determined with more confidence. The spread between the different curves is due to the different crystallite patterns that shift the zero crossing of $dB_y/dy(y)$ away from $y=0$ (where the qubits are located). This shift is up to $20$\,nm, about the size of the assumed crystallite dimensions (not shown).

In view of future devices, we note that surface roughness can be controlled by improving fabrication (line edge roughness $<$ 1 nm are routinely achieved in current semiconductor processing~\cite{LER_IEEE}), reducing its impact on the stray field variation. On the other hand, the large MCA of Co combined with polycrystallinity will still have a detrimental effect on qubit properties, randomly altering $f_{\mathrm{L}}$ and increasing dephasing. One way to reduce stray field variations would be to decrease the MCA by switching to different materials, since $K_1$ varies strongly among ferromagnetic materials and alloys~\cite{cullity2011introduction}. 

\begin{figure*}[ht]
	\centering
	\includegraphics[width=0.95\textwidth]{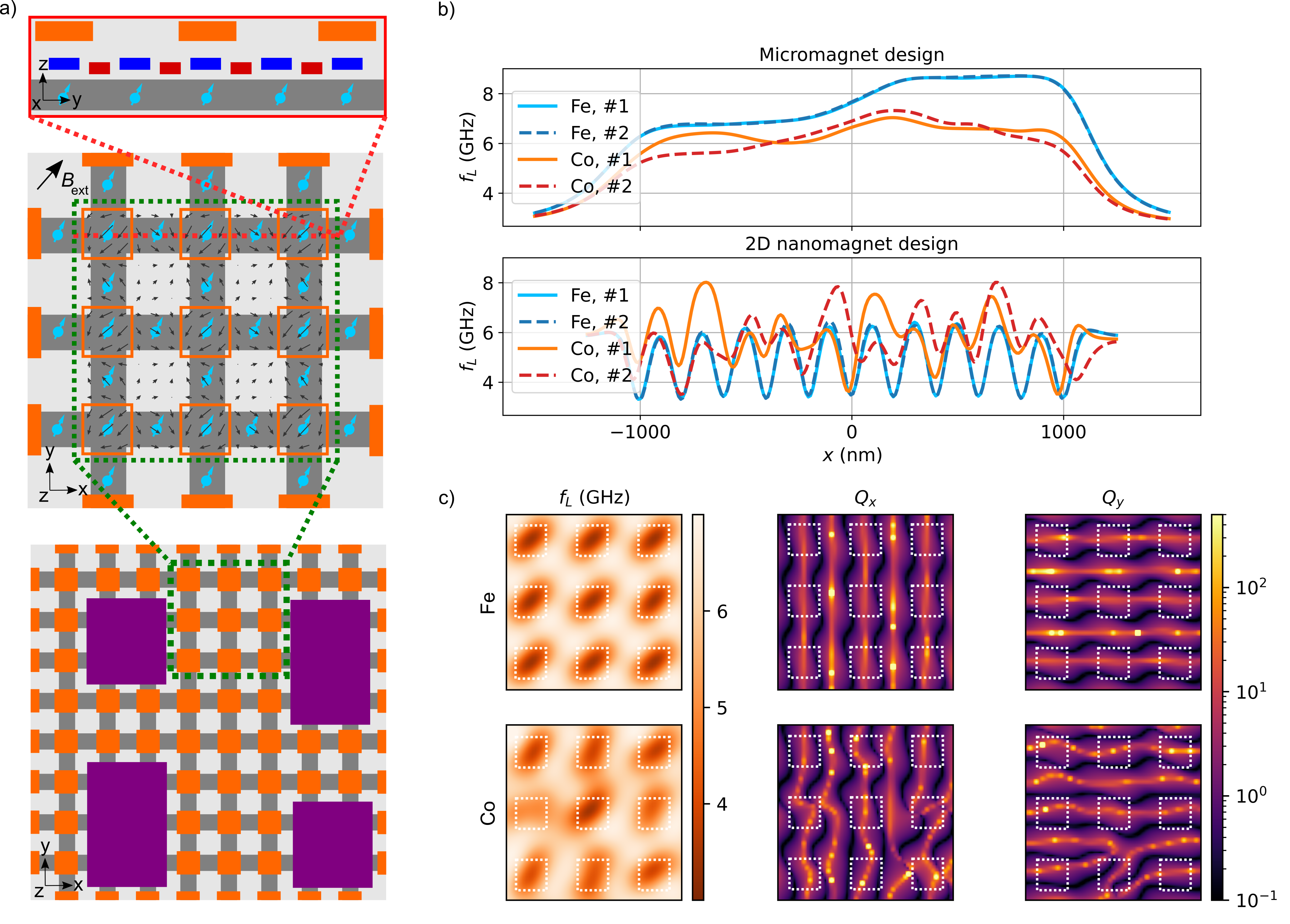}
	\caption{\label{fig:fig_4} (a) Device geometry showing qubits (blue spins) placed on a two dimensional grid (hosting structure schematized by dark grey stripes) and nanomagnets (orange). In the cross-sectional view at the top, plunger gates (dark blue) and barrier gates (red) are also shown. Additional structures (such as charge sensing dots, read-out lines, gate vias) can be placed close to the qubit array into gaps of the qubit arrays (violet rectangles, bottom panel). In the middle panel the black arrows show the in-plane components of the magnetic stray field. (b) Comparison of simulated $f_{\mathrm{L}}$ for qubits placed along varying $x$ positions: micromagnet design discussed in Fig.~\ref{fig:fig_3} (upper panel) and the nanomagnet geometry shown in (a) (lower panel), both for Fe and Co. (c) Simulated $f_{\mathrm{L}}$ and $Q_x$ and $Q_y$ for Fe and Co nanomagnets. The magnetic field deviations are much larger for Co than for Fe. Nanomagnets are squares with edge length of $50$\,nm (contours shown as white dotted lines).}
\end{figure*}

We analyze the effect of material choice by simulating the same magnet design for both Fe and Co in Fig.~\ref{fig:fig_4}(b). We plot $f_{\mathrm{L}}$ along the $x$ direction at $y=0$ and $z=-150$\,nm. The maximum difference between two stray field profiles with two different crystal patterns for Fe is $\approx 0.05$\,GHz, one order of magnitude smaller than for Co, as already seen for the driving gradient. This mirrors the difference of MCA energy among the two materials and makes Fe a better material to drastically decrease the size of the magnets and achieve more uniform magnetization pattern in nanomagnet arrays~\cite{aldeghi2023modular}. 

After having highlighted the relevant parameters of the magnets that impact EDSR manipulation, we now focus our attention on a design that is suitable to drive qubits arranged on a 2D lattice. We suggest a 2D Fe nanomagnet architecture, where the magnets are arranged on a square lattice with a period of twice the qubit spacing (Fig.~\ref{fig:fig_4}). Quantum dots hosting the qubits are placed below and in between each square magnet, tuned by plunger and barrier gates set in planes below the magnet layer. The magnets are either placed above those gates, providing the electric confinement, or themselves act as a gate~\cite{Forster2015, bersano2023quantum, tadokoro2021designs, simion2024qubit} by replacing every second plunger gate, further boosting the attainable driving gradient.

We discuss simulation results with $B_{\mathrm{ext}}=300$\,mT applied in a direction tilted by 45$^\circ$ with respect to the magnet grid. The magnetic stray field oscillates periodically both in its $x$ and $y$ component, as shown in the middle panel of Fig.~\ref{fig:fig_4}(a). This arrangement has several advantages. First, qubit manipulation can be performed by displacing the wavefunction both along the $x$ direction and along the $y$ direction (using the neighboring barrier gates). This opens up the possibility to drive qubits arranged on a 2D grid, either by EDSR or by baseband pulsing~\cite{wang2024}. Second, the magnet arrangement also provides different $f_{\mathrm{L}}$ for neighboring qubits, since the stray field either adds to or subtracts from the external field. The difference between $f_{\mathrm{L}}$ of neighboring qubits is on the order of GHz, substantially larger than the modulation induced by the MCA [Fig.~\ref{fig:fig_4}(b) and (c)]. Moreover, this design is scalable to any qubit grid size without the need for additional magnetic structures to improve single qubit addressability, assuming that the electric field used for the EDSR drive of one qubit does not affect the next-nearest neighbour. The geometry is also flexible in terms of how dense the magnet grid can be made locally. Magnets must be placed at the crossing between vertical and horizontal lines of qubits, but the distance between the next line can be any multiple of twice the qubit pitch. Such flexibility opens up possibilities to accommodate additional structures within the grid spanned by the magnets, such as charge sensing dots, read-out lines or gate vias (Fig.~\ref{fig:fig_4}(a), bottom panel).

We have simulated the nanomagnets as squares with edge length of 50\,nm and thickness of 30\,nm, and plot the simulation at a height of 45\,nm below the lower edge of the magnet. Fig.~\ref{fig:fig_4}(b-c) show that for Fe nanomagnets the stray field pattern repeats with minimal variations along both directions of the magnet array ($\Delta f_{\mathrm{L}}<50$\,MHz). In contrast, for Co nanomagnets $\Delta f_{\mathrm{L}}$ may reach up to 2 GHz. These results highlight the different approach between micro- and nanomagnets. For current EDSR-driven qubits, the global field enabling manipulation and single qubit addressability is modulated by the shape variation of the magnet~\cite{philips2022universal, yoneda2023noise}. Here single qubit EDSR manipulation is performed with a global electric field, meaning that all qubit wavefunctions are displaced simultaneously. Such an approach forces the requirement that all qubits must have different $f_{\mathrm{L}}$ to enable single qubit addressability. This is fulfilled by designing large magnet shape variations with respect to the qubit dimensions, maximizing the driving gradient while keeping the required minimal $\Delta f_{\mathrm{L}}$ between the qubits (Fig.~\ref{fig:SIfig_1}). Fig.~\ref{fig:fig_4}(b) shows that the MCA in Co may strongly modulate the stray field, lifting the engineered $f_{\mathrm{L}}$ differences.

In our architecture, the stray field direction rotates on a length scale commensurate with the qubit pitch. Errors in fabrication and alignment are of similar relevance as for micromagnets (which can be mitigated by improving fabrication), but MCA-induced variations are negligible if the magnets are made of Fe. We also note that the large variation in stray field between neighboring qubits can be advantageous for spin shuttling~\cite{Bosco_shuttling}, thanks to an effective dynamical decoupling during the transfer. The resilience towards stray field variations is also visible in the quality factor plots of $Q_x$ and $Q_y$ [Fig.~\ref{fig:fig_4}c]. We define $Q_i=\frac{dB_z/di}{dB_{\parallel}/di}$, with $i\in (x,y)$, since the qubits have to be displaced either along the $x$ or the $y$ direction depending on their location between the magnets. $B_{\parallel}$ is the magnetic field along the qubit quantization axis. In the Fe simulation, the zero crossing of the dephasing gradient persistently coincides with the maxima or minima of the driving gradient along either the $x$ or $y$ direction, such that $Q_i > 30$ is expected at the foreseen qubit locations.

In conclusion, we have shown that micromagnetic simulations enabled us to accurately model and quantify the impact of sample polycrystallinity, MCA and fabrication-related imperfections on the magnetic stray field. We have compared our simulations with SSM measurements at different external magnetic fields and with experimentally determined spin qubit properties, finding good agreement. Corroborated by these findings, we propose to use Fe as the magnetic material for EDSR magnets because of its higher saturation magnetization and lower MCA than Co. Furthermore, we have developed a qubit architecture with Fe magnets which enables driving qubits on a two-dimensional grid. Our architecture is locally tunable in sparsity and achieves quality factors above 30 with low next-neighbor crosstalk. In addition to demonstrating the importance of micromagnetic simulations to accurately model the performance of nanomagnets for spin qubits, these findings provide a means for designing future architectures for nanomagnet-based spin qubit manipulation.

\begin{acknowledgments}
We thank the Cleanroom Operations Team of the Binnig and Rohrer Nanotechnology Center (BRNC) for advise and support on sample fabrication. We thank Armin Knoll for his support with AFM measurements and Thomas Weber and Marilyne Sousa for their help with XRD measurements. We gratefully acknowledge Brennan Undseth, Lieven Vandersypen and all members of the Spin Qubit team at IBM and of the Poggio Lab for fruitful discussions. This work was supported as a part of NCCR SPIN project funded by the Swiss National Science Foundation (grant number 51NF40-180604). We also acknowledge support of the European Commission under H2020 FET Open grant "FIBsuperProbes" (Grant No. 892427) and the Swiss National Science Foundation under Grant No. 200020-207933.
\end{acknowledgments}

\bibliography{main.bib}

\section{Supplementary Information}

\maketitle
\renewcommand{\figurename}{SI FIG.}
\setcounter{figure}{0}

\subsection{Micromagnetic simulations}
We compute the magnetization pattern with the micromagnetic simulation package MuMax3~\cite{vansteenkiste2014design}. We set the initial magnetization state as a uniform, with the direction parallel to the external field. All the simulations neglect temperature induced magnetic fluctuations, due to the typical low-temperature operation of spin qubits ($<$ 4K) and their marginal effect on dephasing time~\cite{neumann2015simulation, kha2015micromagnets}. The parameters used for each simulation mentioned in the main text are summarized in Table~\ref{tab:table_1}. The choice of parameter follows standard literature values~\cite{cullity2011introduction} for the simulations of Figures~\ref{fig:fig_1} and \ref{fig:fig_4}. For the simulations of the Fe-magnet measured with SSM (Fig.~\ref{fig:fig_3}), we use a thickness of 50 nm, the shape shown in Fig.~\ref{fig:fig_2}(c) and the parameters shown in Table~\ref{tab:table_1}. The saturation magnetization of 1.20\,MA/m was measured by vibrating sample magnetometry on planar films grown in nominally identical conditions, see SI. We take into account MCA by defining randomly shaped and oriented crystallites with average dimension of $40$\,nm with cubic MCA $K_1=300$\,kJ/$\mathrm{m}^3$ (see main text). The size of the crystallites was extracted from AFM, SEM and XRD measurements (see SI). We also corrugate the surface of the magnets by a pattern with a 30\,nm correlation length and a maximum height variation of 20 nm as measured by AFM (see SI). Once the simulation has reached the energy minimum, we convolve the data at a height of 70 nm above the magnet surface with a square filter with an edge size of 110 nm, to compensate for the finite size of the SQUID loop, and rescale the output of the simulation to match the aspect ratio of the cell size in the experiment (38 nm by 67 nm) by spline interpolation. The main axis of the micromagnet has an approximate tilt of 7$^\circ$ with respect to the $y$ direction, which is taken into account in the simulation.

\begin{table*}[ht]
\begin{tabular}{|c|c|c|c|c|c|}
\hline
Figure & Material & Saturation magnetization & Exchange stiffness & MCA & Cell size \\
 &  &  (MA/m) & (pJ/m) &  (kJ/$\mathrm{m}^3$) & (nm)\\
\hline
1 & Co & 1.44 & 30 & none & 10 \\
\hline
2 & Fe & 1.20 & 21 & 300 (cubic) & 10 \\
\hline
3 & Co & 1.11 & 30 & 650 (uniaxial) & 10 \\
\hline
4 & Fe & 1.7 & 21 & 60 (cubic) & 10 (micromagnet)\\
 &  &  & &  &  5 (nanomagnet) \\
\hline
4 & Co & 1.44 & 30 & 650 (uniaxial) & 10 (micromagnet)\\
 &  &  & &  &  5 (nanomagnet) \\
\hline

\end{tabular}
\caption{Parameters for the micromagnetic simulations}

\label{tab:table_1}
\end{table*}

\subsection{Micromagnet designs}

\begin{figure}[h]
	\centering
	\includegraphics[width=0.48\textwidth]{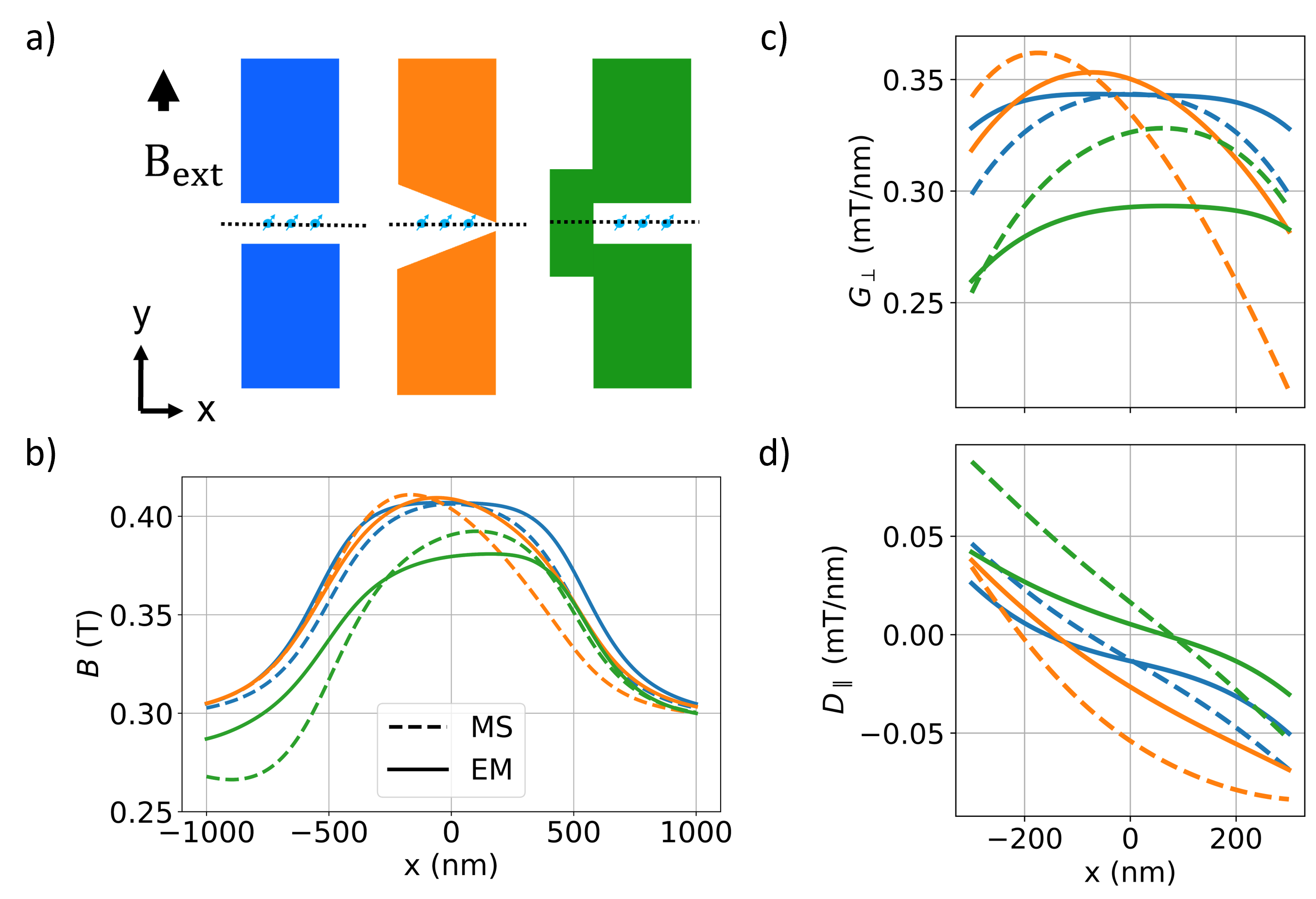}
	\caption{\label{fig:SIfig_1} (a) Schematic device geometry for the designs "Block" (blue), "Tapered" (orange) and "Bridge" (green) (not to scale) (b) Simulated magnetic field ($B$), (c) driving gradient ($G_{\perp}$) and dephasing gradient ($D_{\parallel}$) for the MS (dotted lines) and EM simulation (continuous lines) of the three designs plotted with the same color code as the sketches in (a).}
\end{figure}

The micromagnet design, the external field direction and the qubit array orientation define which field components of $B$ should be used for EDSR drive or will contribute to the qubit dephasing. The geometries shown in SI Fig.~\ref{fig:SIfig_1}(a) set the EDSR driving to use the slanting magnetic field coming from the magnets along the out-of-plane direction ($z$) and displace the electron wavefunction along the external magnetic field $B_{\mathrm{ext}}$ ($y$), which is parallel to the longest axis of the magnets and perpendicular to the qubit chain ($x$). The spatial arrangement of magnets, $B_{\mathrm{ext}}$ direction and qubit positions sets: i) $f_{\mathrm{L}}$ should vary along the $x$ direction, ii) the driving gradient is $G_{\perp}= \frac{dB_z}{dy}+\frac{dB_x}{dy}$ and iii) the dephasing gradient is $D_{\parallel} = \frac{dB_y}{dx} + \frac{dB_y}{dy}$. Displacement along $z$ is neglected because we expect strongly suppressed displacements along this direction for qubits confined in semiconductor heterostructures. 

The dephasing gradient is linked to the dephasing rate $\Gamma$ by the charge noise induced wavefunction displacement $\Delta_i$ along the $i$ direction, according to the equation~\cite{aldeghi2023modular}:
$\Gamma=\sum_i\Gamma_i=\sum_i\pi\sqrt{2}\frac{\gamma_e}{h}\frac{dB_{y}}{di}\Delta i$ where $\gamma_e$ is the electron gyromagnetic ratio (g-factor of 2), $h$ Planck's constant and $dB_{y}/di$ the first derivative of the stray field along the $i$ direction.
This displacement is induced by charge noise affecting the local confinement potential~\cite{connors2019low}. Here, we assume that such displacements are identical in amplitude along the two in-plane directions $x$ and $y$, but strongly suppressed along $z$ because of the considerably stronger confinement potential along the growth direction, such that $\Gamma = \Gamma_x+\Gamma_y =\pi\sqrt{2}\frac{\gamma_e}{h}(\frac{dB_{y}}{dx}+\frac{dB_{y}}{dy})\Delta$. 

We now compare three possible micromagnet designs: "Block", "Bridge"~\cite{Yoneda_robust} and "Tapered"~\cite{Brunner11}. The magnets have width $W= 1000$\,nm, length $L= 3000$\,nm, thickness $T= 200$\,nm and a variable gap $G$. The tapering angle is 8.5$^\circ$, while the additional block for the design "Bridge" has a width $W_b= 400$\,nm and a length $L_b= 1200$\,nm. We assume the qubits to be lined up along the $x$ direction at $y=0$ and $z=-200$\,nm (100 nm below the edge of the magnets) and plot $B$ along this line [SI Fig.~\ref{fig:SIfig_1}(a)].

We discuss first the MS simulation and how single qubit addressability is achieved in these magnet geometries. The first design is based on two blocks with a constant gap in between [as already introduced in Fig.~\ref{fig:fig_1}(a)], and creates a bell-shaped profile along the foreseen qubit chain position, symmetrical with respect to $x= 0$\,nm. Single qubit addressability is achieved by setting a large enough spread of the qubit frequencies with respect to the Rabi frequency ($\Delta f_{\mathrm{L}} > 2f_{\mathrm{Rabi}}$)~\cite{Yoneda_robust}, which in this case may be achieved by placing the qubits only along $x > 0$ ($<0$), such that $f_{\mathrm{L}}$ would monotonically decrease (increase) along the chain [SI Fig.~\ref{fig:SIfig_1}(b)]. 

In the design "Tapered" the field lines are pushed outside the gap between the magnets depending on the gap size, such that larger fields are found below the region where the gap is wider [SI Fig.~\ref{fig:SIfig_1}(b)]. We note that this trend is the opposite from what is found at height $z=0$.  

The design "Bridge" can be decomposed into 3 magnetic dipoles, with two aligned along their longest axis and one smaller between them and displaced to the side (i.e. the additional block bridging the magnets in the "Block" geometry). The field lines connect the two opposite magnetic poles across the gap along the direction of $B_{\mathrm{ext}}$, similarly to the two designs before. Likewise, also the smaller dipole on the side provides a magnetic field, but its field lines have opposite direction at the qubit location. This, combined with the stray field decay along the $x$ direction [analogous to what was shown for the $z$ direction in Fig.~\ref{fig:fig_1}(d)], reduces the total field depending on the $x$ location, creating the desired difference in $f_{\mathrm{L}}$ between the qubits [SI Fig.~\ref{fig:SIfig_1}(b)]. We note that this strategy to modulate the stray field is the most successful in keeping a constant driving gradient and single qubit addressability along the $x$ direction [SI Fig.~\ref{fig:SIfig_1}(c)], in agreement with~\cite{Yoneda_robust}.  

The aforementioned considerations about single qubit addressability hold also for the EM simulation, with the important difference that the magnetization pattern relaxation lifts the symmetry of the fields assumed in the MS approximation. In the design "Block", we see that the stray field profile is flattened, due to the rotation of the magnetic field along the edge of the magnet. As forced by the vortex formation, the magnetization at $x=0$ would point along the $x$ direction, but the stray field minimization rotates it along the $z$ direction [Fig.~\ref{fig:fig_1}(b)]. The balance between these two energies changes along the edge of the magnet, such that the rotation does not happen monotonically along the $xz$ plane facing the qubits, leading to the almost flat profile of $f_{\mathrm{L}}$.

For the design "Tapered", the relaxed magnetization pattern results in a slightly smaller $B$ peak that is shifted along $-x$. Also in this case the $f_{\mathrm{L}}$ difference between qubits would be smaller than expected, such that qubits placed symmetrically around the peak maxima may have identical $f_{\mathrm{L}}$. The same result is found in the design "Bridge", where the formation of a vortex along the surfaces in the $xz$ planes reduces the differences in $f_{\mathrm{L}}$ between neighboring qubits. For all three designs the single qubit addressability is reduced, impacting significantly the crosstalk between them. Similarly, also the dephasing and driving gradient show significant differences between the MS and EM approach. These results highlight that the relaxation of the magnetization pattern may jeopardize the expected ESDR drive of the qubits predicted by MS simulations.

\subsection{Edge rounding effect}

\begin{figure}[ht]
	\centering
	\includegraphics[width=0.48\textwidth]{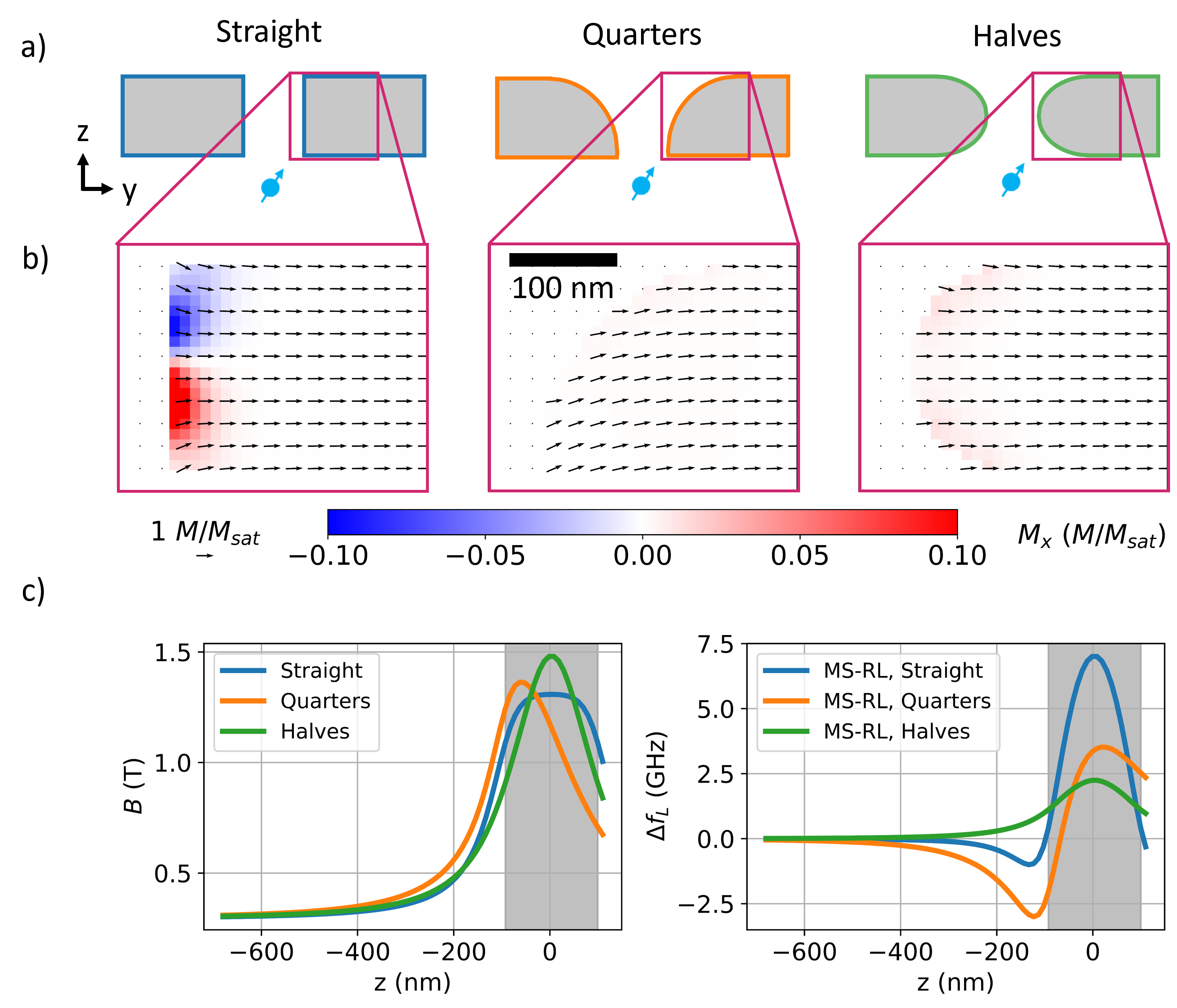}
	\caption{\label{fig:SIfig_2} Effect of the shape of the cross section on the stray field profile. (a) device geometry. (b) Magnetization pattern at $x$=0. In-plane components are shown as arrows, the out-of-plane component as color. (c) EM simulation of the magnetic field ($B$) (left panel) and $\Delta f_{\mathrm{L}}$ (right panel) between the MS and the EM simulations for different heights. The grey shaded areas show the position of the magnets.}
\end{figure}

Here we discuss the role of non-ideal edges of the magnetic structures, as given e.g. by lithographic processes. We use the same block geometry as in Fig.~\ref{fig:fig_1} with $G=100$\,nm and $T= 200$\,nm (named here "Straight"), but we round the highlighted surfaces such that the $y-z$ profile is a quarter of a circle ("Quarters") or a semicircle ("Halves"). Due to the rounded shape, for both simulations higher $B$ maxima are found with respect to the "Straight" simulation, since the magnetization pattern does almost not rotate into the $x$ direction [SI Fig.~\ref{fig:SIfig_2}(b)]. Indeed, the "Halves" shape reaches the largest stray field value in the volume between the magnets, since its magnetization pattern is almost fully aligned with the external field. For the "Quarters" shape, the induced sharp edge forces the magnetization at the surface to rotate towards the $z$ direction, following the curvature of the rounded surface. Because of this, the stray field decreases within the gap at $z=0$ but increases significantly for small negative $z$ values [SI Fig.~\ref{fig:SIfig_2}(c), right panel]. For the half-circle shape, the magnetization pattern keeps the inversion symmetry at $z=0$, but due to less magnetic material at the edges the "Halves" shape provides smaller fields than the "Straight" one already at $z= -42$\,nm. Rounded edges therefore impact the magnetization pattern significantly, with differences in $f_{\mathrm{L}}$ of up to few GHz in comparison to straight edged magnets (e.g. the $f_{\mathrm{L}}$ difference between the "Straight" and "Quarters" simulation at $z=-120$\,nm is $4.08$\,GHz).

\subsection{SSM experiment}
\subsubsection{SQUID-on-lever fabrication}

We fabricate a nano SQUID via Ga-FIB milling on a Nb covered AFM cantilever, as detailed in~\cite{wyss2022magnetic}.
The sensor is characterized and operated at 4.2~K in a semi-voltage biased circuit, with its current response $I_{\text{SQUID}}$ measured by a series SQUID array amplifier (Magnicon). The SQUID loop has an effective diameter of 110~nm, determined from its quantum interference pattern. 

\subsubsection{Magnetic imaging}

Magnetic imaging is conducted using a custom-built scanning probe microscope under high vacuum within a \(^4\text{He}\) cryostat. Since $I_{\text{SQUID}}$ modulates with the magnetic flux, this response provides a measure of the local magnetic field threading through the SQUID loop. The axis of the SQUID is tilted by $10^\circ$ relative to the $z$ direction with the tilt occurring in the $x$ direction. This ensures that the cantilever apex (where the SQUID loop is located) can approach the sample surface without the cantilever body making contact. We scan the sample using a scanning probe controller (Specs) and piezoelectric actuators (Attocube) at a constant SQUID-sample spacing of 70~nm, measured from the magnet upper surface. The spatial resolution is given by this spacing and the effective diameter of the SQUID loop of 110~nm. During imaging, an out-of-plane magnetic field ($B^{\mathrm{ext}}_z$) between 5 and 50~mT is applied to operate the SQUID near an inflection point in its interference pattern. The complete list of applied fields during the scans is
shown in Table~\ref{tab:table_2}.

\subsubsection{SQUID calibration}
The SQUID is primarily sensitive to out-of-plane magnetic fields. 
However, due to the $10^\circ$ tilt of the SQUID axis relative to the z-axis, a small component of the in-plane magnetic field along the y-axis can also thread through the SQUID loop. 
In addition, strong in-plane magnetic fields reduce the density of superconducting charge carriers in the SQUID, suppressing its current modulation. 
Both effects make the SQUID sensitive to in-plane fields, although the primary sensitivity to out-of-plane fields remains. To account for these effects, we measure the response of $I_\text{SQUID}$ as a function of both out-of-plane applied magnetic field $B^{\mathrm{ext}}_z$ and in-plane applied magnetic field $B^{\mathrm{ext}}_y$ after each scan, over a range of $\pm150$~mT around the applied fields used during the scans. 
Before each scan, the Fe nanomagnets are initialized by applying $B^{\mathrm{ext}}_y=800$~mT, followed by sweeping down to the field used during the scan. 
Since $I_\text{SQUID}$ is sensitive to both in-plane and out-of-plane fields ($I_\text{SQUID}(B_z, B_y))$, we have to determine the stray field of the nanomagnet iteratively. \\

Initially, we assume that in-plane field components do not affect $I_\text{SQUID}$. 
We hence set $B_y = 0$ at every position and solve for an initial $B_z$ using $B_z(I_\text{SQUID},B_y = 0)$ and plugging in the measured $I_\text{SQUID}$. Following $\nabla \cdot \mathbf{B} = 0$ and Maxwell's equations~\cite{Broadway2020}, we use this initial $B_z(x,y)$ to solve for $B_y(x,y)$. We can then obtain a refined map of $B_z(x,y)$ by solving for $B_z(I_\text{SQUID}, B_y)$ at each position and plugging in the measured $I_\text{SQUID}$ and the new values of $B_y$. We iterate in this way until the difference between successive $B_z (x, y)$ maps becomes negligible.

Although an out-of-plane field is applied to operate the SQUID in the linear response region, the stray magnetic field from the nanomagnets exceeds 100~mT, pushing the calibration beyond the half-period range of the interference pattern (i.e., the region between a maximum and a minimum). Consequently, \(I_\text{SQUID}(B_z)\) becomes multi-valued, complicating the calibration process.   
To address this, the multi-valued modulation curve \(I_\text{SQUID}-B_z\) is cropped at the extrema to create out-of-plane field ranges with a monotonic response. Each pixel in the scan is assigned to a specific field region corresponding to a monotonic calibration curve \(I_\text{SQUID}-B_z\).
However, near the maxima and minima of \(I_\text{SQUID}(B_z)\), the signal-to-noise ratio is low, resulting in noisy regions with large errors. These errors propagate to the $B_x (x, y)$ and $B_y (x, y)$ components, and the iterative calibration process cannot compensate for them. Additionally, regions with a strong $B_x$ component also exhibit larger errors, as the calibration only considers $B_z$ and $B_y$ components, without accounting for the reduction in $I_{\text{SQUID}}$ caused by strong $B_x$ fields. 

\begin{table}[h]
\begin{tabular}{|l|c|}
\hline
Scan name & $\bm{B_{\mathrm{ext}}}=(B_x, B_y, B_z)$ (mT) \\
\hline
0.5 & (500,0, -3) \\
\hline
0.38 & (380, 0, -13) \\
\hline
0.1 &  (100, 0, -45) \\
\hline
0 & (0, 0, -43) \\
\hline
-0.1  & (-100, 0, -53) \\
\hline
-0.38 & (-380, 0, 27) \\
\hline
\end{tabular}
\caption{External magnetic field applied to the micromagnet during the SSM scans.}

\label{tab:table_2}
\end{table}

\subsection{Saturation magnetization measurement}
\begin{figure}[ht]
	\centering
	\includegraphics[width=0.48\textwidth]{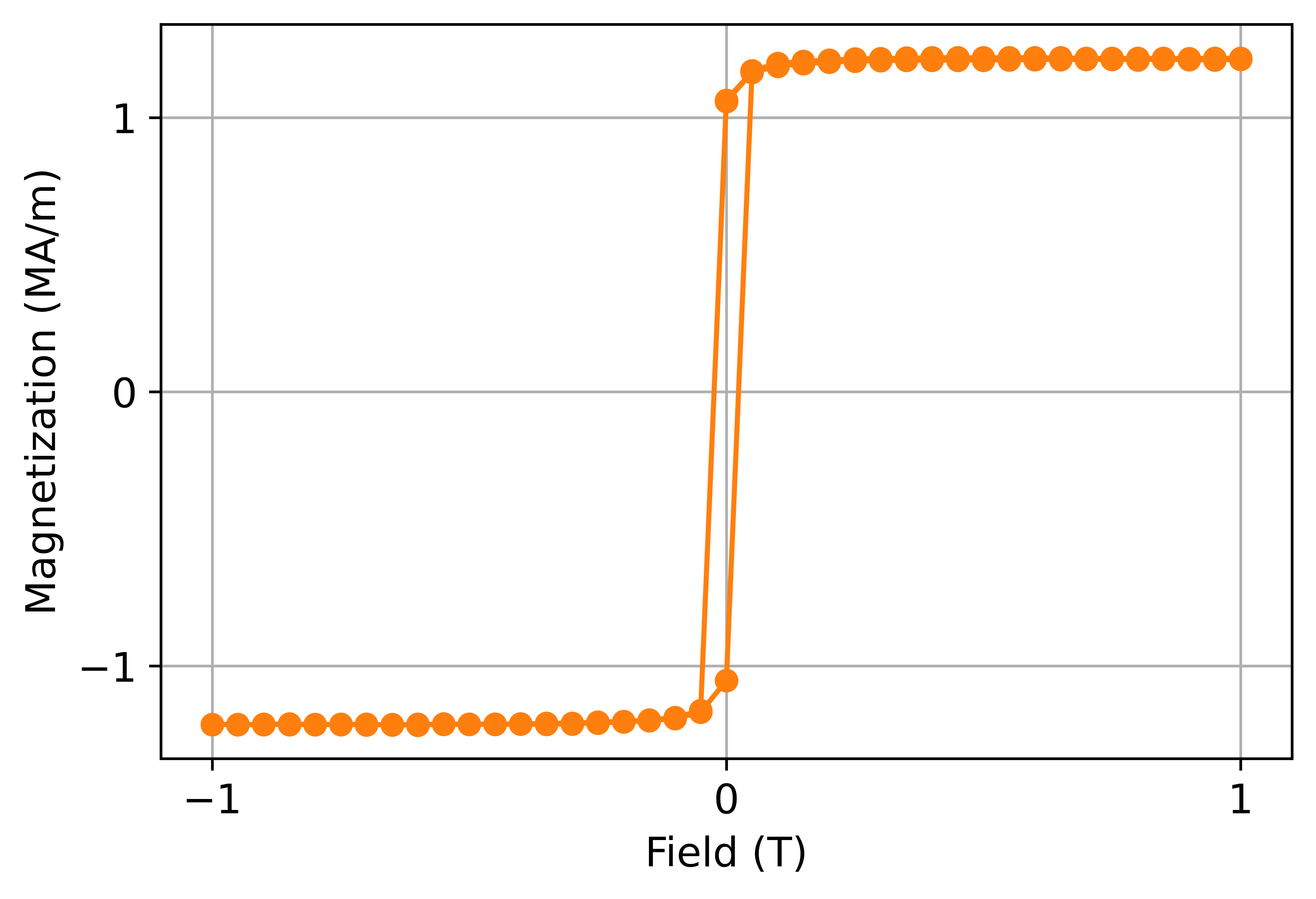}
	\caption{\label{fig:SIfig_4} Vibrating sample magnetometry on a blanket film of Fe. We found $M_{\mathrm{sat}}=1.2$\,MA/m, which correspond to a 30\% reduction from the literature value of 1.7\,MA/m~\cite{cullity2011introduction}.}
\end{figure}

We use a Lake Shore Cryotronics 7300 Series Vibrating Sample Magnetometer to determine the saturation magnetization. The sample used is a 10x10 mm$\mathrm{^2}$ square and 50 nm thick blanket Fe film deposited by e-beam evaporation, under the nominally same conditions as for the growth of the micromagnet in Fig.~\ref{fig:fig_2}. We measure an hysteresis loop by applying the external field parallel to the sample surface (i.e. within the easy plane set by shape anisotropy). We calculate the magnetization under the assumption that the whole 50 nm deposited film is magnetic, finding $M_{\mathrm{sat}}=1.2$\,MA/m (SI Fig.~\ref{fig:SIfig_4}).

\subsection{Crystallite size and texture determination}
\begin{figure*}[ht]
	\centering
	\includegraphics[width=0.95\textwidth]{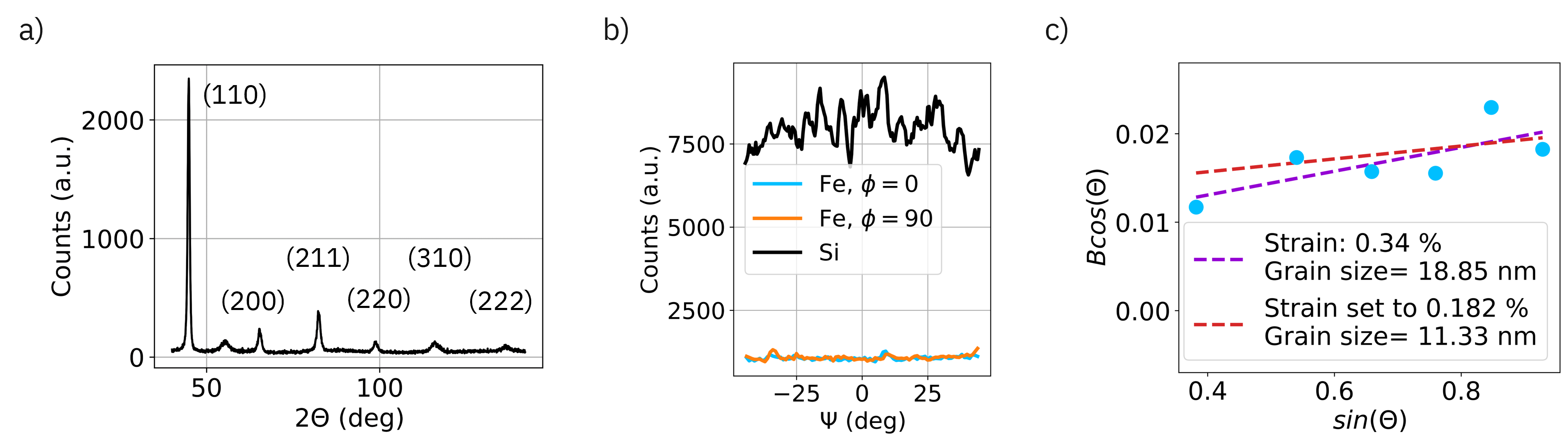}
	\caption{\label{fig:SIfig_5} (a) Grazing incidence XRD measurement, with the crystalline plane index reported next to the peak. (b) Rocking curve measurement along $\Psi$, such as to keep the grazing incidence angle scan constant. The curve labeled Si belongs to a Si polycristalline sample with known random texture. (c) Williamson-Hall plot with the peaks FWHM extracted from the measurement shown in Fig.~\ref{fig:SIfig_5}(a). The dashed curve in violet is the best fit for the data points. The red dashed curve is the best fit under the assumption of a uniform strain of 0.182\% throughout the film. }
\end{figure*}

We investigated the crystallite size and texture of a blanket Fe film with 50 nm thickness grown in nominally identical conditions as the magnet scanned in Fig.~\ref{fig:fig_2}, since standard XRD can be performed only on structures considerably larger than the micromagnets. We use a Panalytical X'Pert MRD tool in grazing incidence (grazing angle is 1$^\circ$). The planar film is polycrystalline and shows no texture, as confirmed by the XRD scan and a rocking curve performed along the incident beam direction (SI Fig.~\ref{fig:SIfig_5}). We plot the full-width at half maxima (FWHM) of the peaks angle dependence after correcting for the instrumental broadening in the Williamson-Hall plot. The scattering of the data points is considerably larger than the uncertainties due to the peak fitting, invalidating the assumptions used for extracting strain and crystallite size in the conventional Williamson-Hall analysis. Such deviations are caused by anisotropic crystallites, as visible in the SEM figure [SI Fig.~\ref{fig:SIfig_6}(a)]. Under the drastic assumption of uniform strain within the film, we can still try to extract a lower boundary for the crystallite size. We roughly estimate the strain $\epsilon$ using the formula $\epsilon= (\alpha_{\mathrm{Fe}}-\alpha_{\mathrm{Si}})\Delta T$, where $\alpha_{\mathrm{Fe, Si}}$ is the coefficient of thermal expansion of Fe respectively Si and $\Delta T$ the temperature difference between measurement temperature and the deposition temperature, which we estimate to be 200 K. We calculate then a strain of 0.182\%, and under this strain assumption we find that a value of approximately 11 nm for the crystallite size fits best. This crystallite size is in agreement with the smallest side of the elongated crystallites measured in the SEM images.

\subsection{Surface roughness measurement}
\begin{figure*}[ht]
	\centering
	\includegraphics[width=0.95\textwidth]{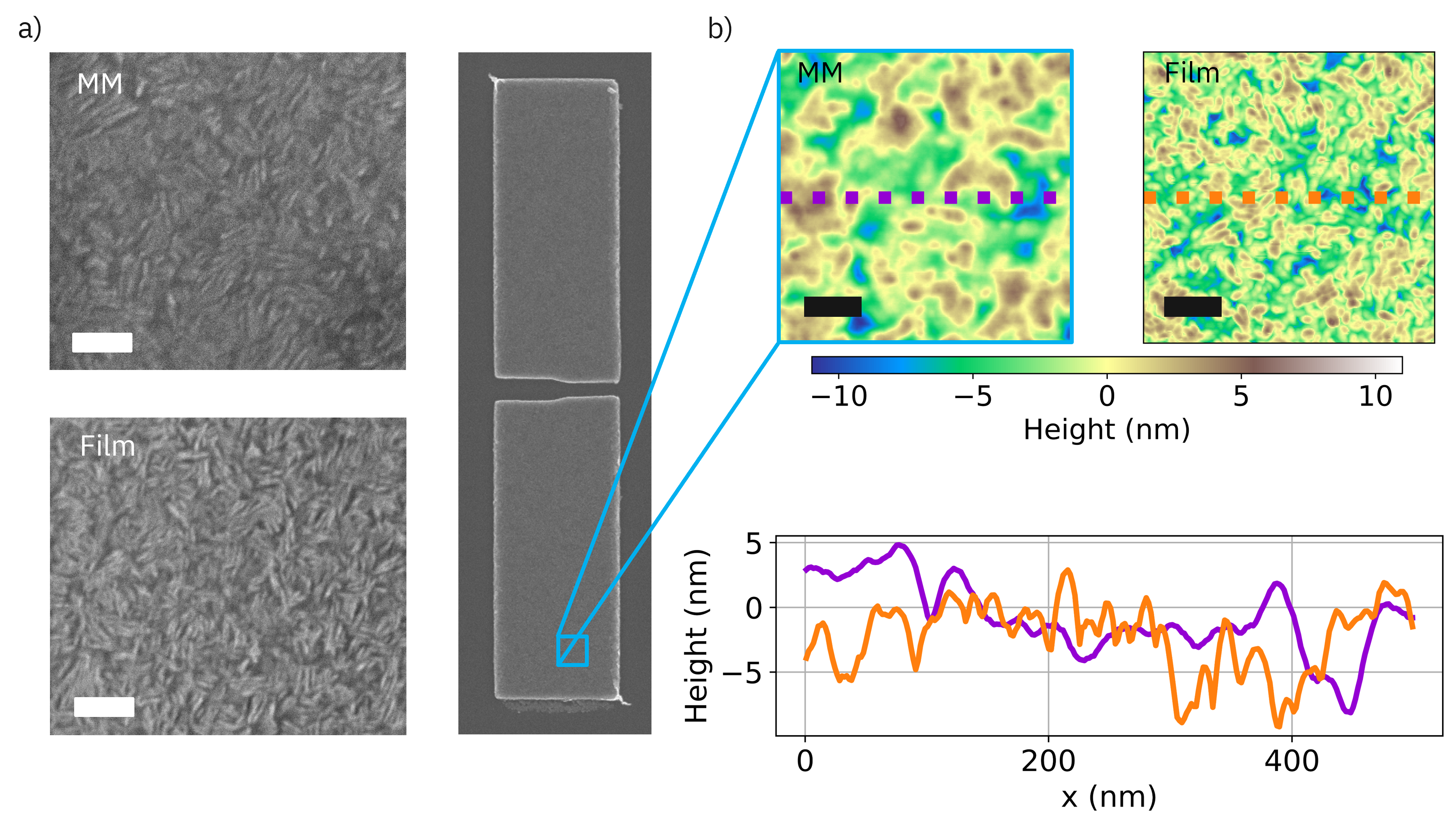}
	\caption{\label{fig:SIfig_6} (a) SEM images of the micromagnet scanned in Fig.~\ref{fig:fig_3} ("MM") and a planar Fe film grown in nominally identical conditions ("Film"). (b) AFM measurement of the same magnets, where the panel below reports the data along the cross section shown. All scale bars are 100\,nm.}
\end{figure*}

We performed AFM measurements (Park Systems, NX20) on the micromagnet scanned in Fig.~\ref{fig:fig_2}, finding a correlation length of 32.34\,nm and a RMS roughness of 1.14\,nm (SI Fig.~\ref{fig:SIfig_6}). Immediately afterwards and with the same AFM tip we scanned a 50\,nm thick planar Fe film grown in nominally identical conditions as the micromagnet. For the planar film the correlation length is 13.76\,nm and the RMS roughness is 1.28\,nm. The surface pattern visible in the SEM picture is similar, but the correlation length differs. We ascribe this difference to remaining traces of resist after the development following the electron beam lithography. Based on the SEM findings, we see in both images that grains are elongated, with a short side of approximately 10\,nm and a long side ranging between 30 and 50\,nm. We then assume a crystallite size of 40 nm for the simulations described in the main text. 

\subsection{Gradient extraction}
\begin{figure*}[ht]
	\centering
	\includegraphics[width=0.7\textwidth]{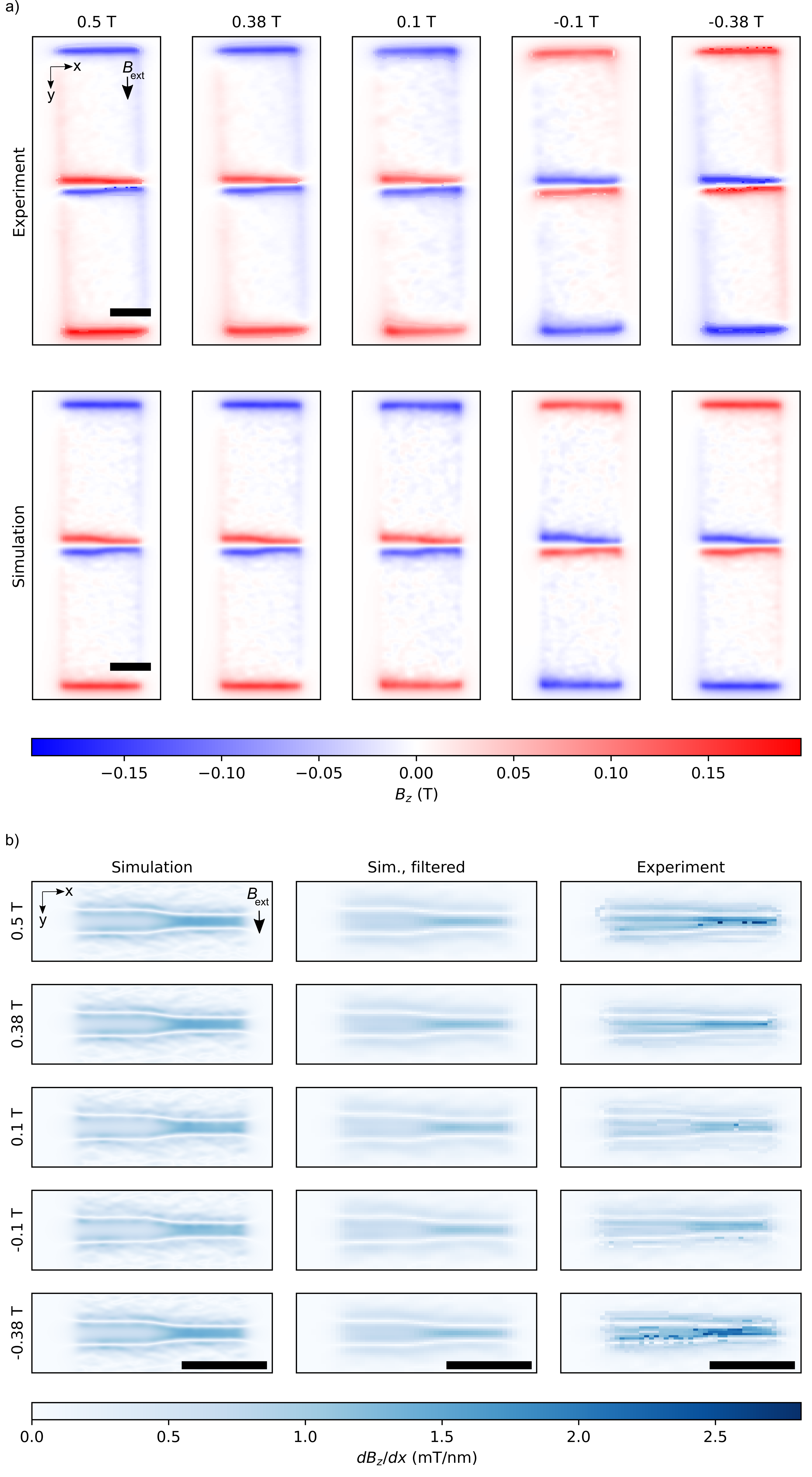}
	\caption{\label{fig:SIfig_7} (a) SSM scans of $B_z(x,y)$ at different external magnetic fields (upper panel) and corresponding micromagnetic simulations (lower panel). (b) Numerical derivative $dB_z/dx$ in the region of the gap in (a); the first column shows the unprocessed simulation, the second column depicts the data of the first column smoothed with a uniform filter with 110 nm in size as to mimic the averaging performed by the SQUID loop, and the last column shows the processed SSM measurements at different $B_{\mathrm{ext}}$. The scale bars are 1000 nm.}
\end{figure*}

As explained in the main text, the micromagnets are designed to provide a large field gradient at the qubit location to perform EDSR drive. During spin qubit experiments, it is challenging to quantify the exact magnetic field gradient, since the experimentally measurable Rabi frequency ($f_{\mathrm{Rabi}}$) is weighted also by the wavefunction displacement ($\Delta$)~\cite{Kawakami2014}: $f_{\mathrm{Rabi}}=\frac{g\mu_BG_{\perp}}{h}\d=\Delta$. SSM provides a direct method to measure the driving gradient $G_{\perp}$. In this geometry $G_{\perp}=dB_z/dy$, such that the gradient can be extracted by taking the numerical derivative along the $y$ direction of $B_z$. The result of this simple approach is shown in SI Fig.~\ref{fig:SIfig_7}. Unfortunately, the experimental data show a high noise level in the gap region, especially at $\lvert B_{\mathrm{ext}} \rvert > 100$\,mT [SI Fig.~\ref{fig:SIfig_7}(a)]. In the attempt to remove the experimental noise, we fit a 8th degree polynomial along the $y$ direction to each linescan between $y=\pm 1000$\,nm [one representative dataset is shown in Fig.~\ref{fig:fig_2}(c)]. The results of the fit are shown in Fig.~\ref{fig:fig_2}(e). SI Fig.~\ref{fig:SIfig_7}(a) shows an overview of all performed scans under the conditions reported in Table~\ref{tab:table_2}, and Fig.~\ref{fig:SIfig_7}(b) the numerical derivative of these scans.

\subsection{Variable gap magnet design}
\begin{figure*}[ht]
	\centering
	\includegraphics[width=0.95\textwidth]{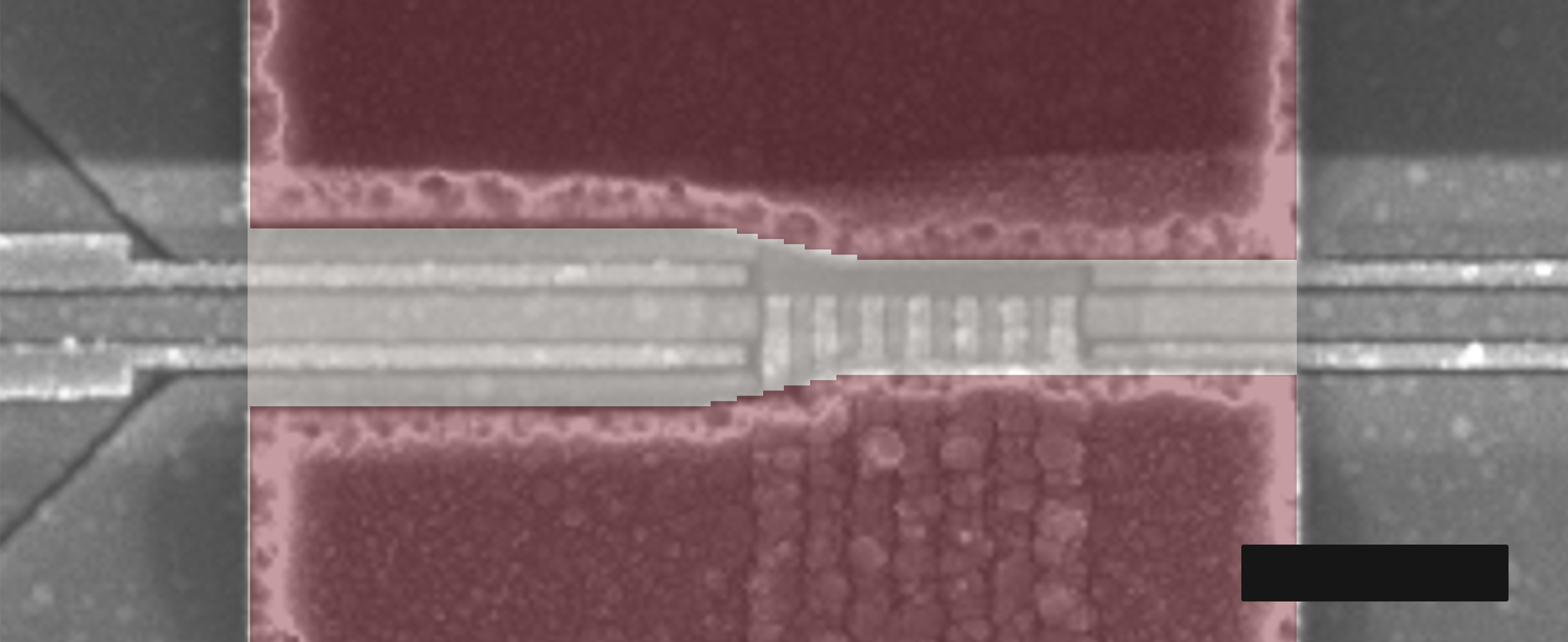}
	\caption{\label{fig:SIfig_8} Overlay of our simulated magnet design (red) with an SEM image of a sample made in the same fabrication run as the one described in~\cite{philips2022universal} (greyscale). The scale bar is 500 nm. SEM image courtesy of Lieven Vandersypen (TU Delft) and TNO.}
\end{figure*}

We mimic the magnet design of Philips et al.~\cite{philips2022universal} in our simulations by setting a width $W= 1000$\,nm, length $L= 5000$\,nm and thickness $T= 200$\,nm, and approximate the triangular cross-section of the magnet edges as a staircase with a height of 200\,nm, a base of 100\,nm and the step edge matching the cell size of 10\,nm. We also take into account a 40\,nm misalignment along the $x$ direction between the two magnets. The gaps are set to 280\,nm and 430\,nm. We show an overlayed image of our simulation design and a SEM image of a sample made in the same fabrication run as the one used in the experiment in~\cite{philips2022universal} in SI Fig.~\ref{fig:SIfig_8}. From the SEM image a granular structure with features size of approximately 40 nm is visible. We thus use a crystallite size of 40 nm in the simulations described in the main text.

\end{document}